\documentclass{vgtc}                          

\ifpdf
  \pdfoutput=1\relax                   
  \usepackage{graphicx}                
  \DeclareGraphicsExtensions{.pdf,.png,.jpg,.jpeg} 
\else
  \usepackage{graphicx}                
  \DeclareGraphicsExtensions{.eps}     
\fi%

\graphicspath{{figures/}{pictures/}{images/}{./}} 

\usepackage{microtype}                 
\PassOptionsToPackage{warn}{textcomp}  
\usepackage{textcomp}                  
\usepackage{mathptmx}                  
\usepackage{times}                     
\usepackage{cite}                      
\usepackage{tabu}                      
\usepackage{booktabs}                  

\usepackage{cite}
\usepackage{amsmath,amssymb,amsfonts}

\usepackage{algorithm}
\usepackage[noend]{algpseudocode}
\usepackage{graphicx}
\usepackage{textcomp}

\usepackage{subfiles}
\usepackage{kotex}
\usepackage{enumitem}

\newcommand{\githruext}{ExIF}
\newcommand{\parahead}[1]{{\vspace{0.05cm}\fontfamily{phv}\selectfont{#1} \ \ }}

\onlineid{1032}

\vgtccategory{Application/design study}

\vgtcinsertpkg

\title{Interactive Visualization for Exploring Information Fragments \\in Software Repositories}

\author{
    Youngtaek Kim\textsuperscript{1 4} 
    \quad
    Hyeon Jeon\textsuperscript{1}
    \quad
    Kiroong Choe\textsuperscript{1}
    \thanks{e-mail: \{ytaek.kim, hj, krchoe\}@hcil.snu.ac.kr}
    \quad
    Hyunjoo Song\textsuperscript{2}
    \thanks{e-mail: hsong@ssu.ac.kr}
    \quad
    Bohyoung Kim\textsuperscript{3}
    \thanks{e-mail: bkim@hufs.ac.kr}
    \quad
    Jinwook Seo\textsuperscript{1}
    \thanks{e-mail: jseo@snu.ac.kr  \vspace{-55pt}}
}

\affiliation{
    \vspace{-12pt}
    \scriptsize \textsuperscript{1} Seoul National University, Seoul, Republic of Korea
    \quad
    \scriptsize \textsuperscript{2} Soongsil University, Seoul, Republic of Korea 
    \\
    \scriptsize \textsuperscript{3} Hankuk University of Foreign Studies
    \quad
    \scriptsize \textsuperscript{4} Samsung Electronics, Suwon-si, Gyunggi-do, Republic of Korea\\
} 

\teaser{
 \vspace{-8pt}
 \centering
 \includegraphics[width=\textwidth]{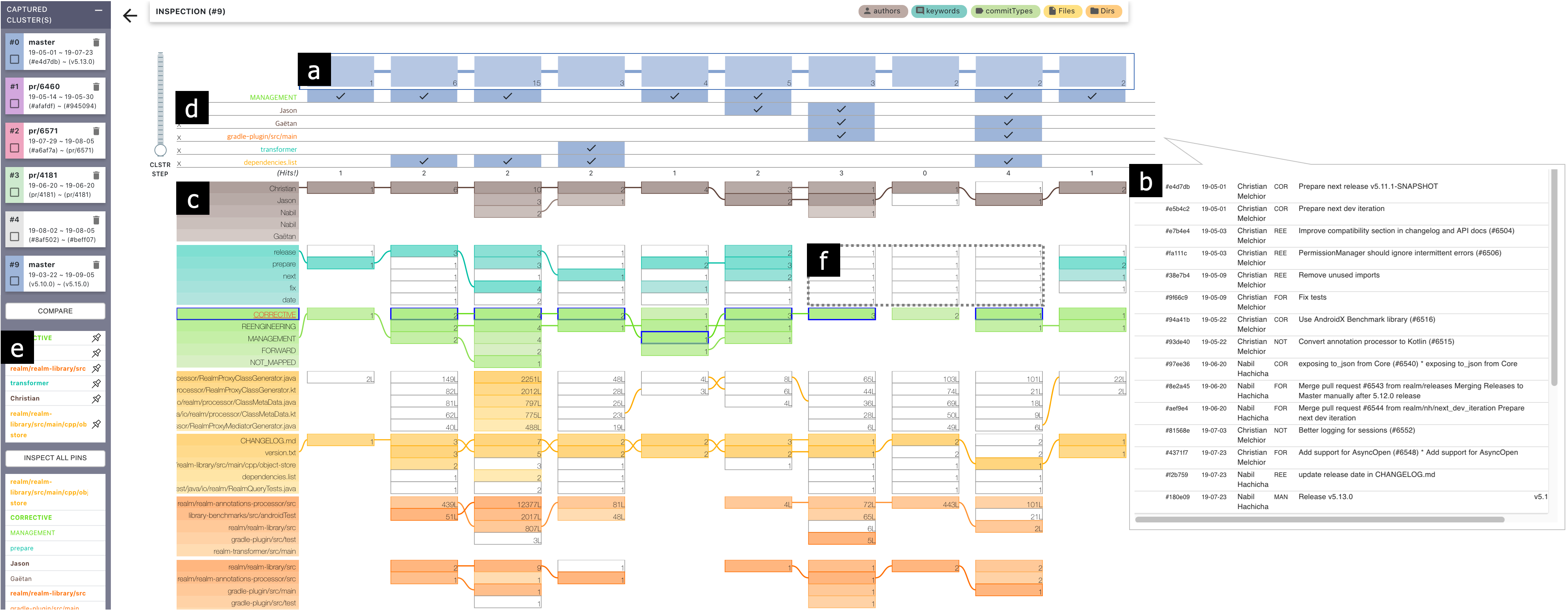}
 \caption{
    Interactive visualization for exploring information fragments. (a) Clusters in the selection scope are displayed. (b) Clicking on a cluster box will display the details of the cluster. (c) The dimension value table provides an overview and details of clusters. (d) One can examine if each cluster contains the selected fragments through inspection view. (e) Fragment view presents the history of fragments with pinning. (f) The clusters in the dashed box did not include common values for the keyword dimension.
 }
 \label{fig:teaser}
}

\abstract{

Software developers explore and inspect software repository data to obtain detailed information archived in the development history.
However, developers who are not acquainted with the development context suffer from delving into the repositories with a handful of information; they have difficulty discovering and expanding information fragments considering the topological and sequential multi-dimensional structure of repositories.
We introduce \githruext{}, an interactive visualization for exploring information fragments in software repositories. \githruext{} helps users discover new information fragments within clusters or topological neighbors and identify revisions incorporating user-collected fragments.

} 

\CCScatlist{
  \CCScatTwelve{Human-centered computing}{Visu\-al\-iza\-tion}{Visualization systems and tools}{Visualization toolkits};
}

\begin{document}

\renewcommand{\figureautorefname}{Figure}


\firstsection{Introduction}

\maketitle

Developers explore software repository data to understand development history. 
A developer who has joined a new project or needs to find a solution related to a historical issue can acquire information from the repository that archives sets of changed files and metadata such as author, message, date, and branch name. 

If developers are not acquainted with the development context or not aware of important context, whether entirely or partially, they need to discover enough information fragments (subsets of development information) \cite{fritz2010using} from the repository. 
However, they can suffer from exploring the repository due to the inherent complex topology and dimensionality of the metadata.
In the case of Git repository exploration, one can discover a new information fragment from clusters of commits containing similar changes or topologically neighboring commits; for example, they can consider sequential changes because commits preserve only differences between commits; or they can query keywords along an appropriate dimension for finding important commits.

Developers mostly use GitHub or their own scripts using the \texttt{git-log} for navigating and searching information fragments. 
RepoVis \cite{feiner2018repovis} supports a full-text search for repositories with a comprehensive visual overview, and Cosentino et al. \cite{cosentino2015gitana} proposed a Git repository inspector by SQL querying based on the conceptual schema. 
However, such works focused on supporting keyword search rather than on discovering and expanding information fragments.

Meanwhile, similar issues have been addressed in the literature search field as the underlying structure of academic metadata is also huge and complicated, like git metadata. Sultanum et al. \cite{sultanum2020understanding} conducted an interview with researchers to draw design goals throughout the entire literature review workflow. 
Choe et al. \cite{choe2021papers101} investigated further into the content discovery and inclusion stage in the workflow.
They distilled finer user needs such as discovering and expanding from the first seed paper or seeking serendipity to find relevant papers. Although these works are from different fields, we could adopt the discovery process and task objectives from them.


In this paper, we introduce \githruext{}\footnote{\vspace{-20pt}The demo is available at https://githru.github.io/exif/.}, an interactive visualization for exploring information fragments in software repositories.
\githruext{} is implemented based on the features of the Githru \cite{kim2020githru} system to provide a scalable overview of a Git repository's complex topology.
The proposed visualization supports discovering and expanding information fragments from clusters or topological neighbors through efficient navigation of the topology. 
Also, \githruext{} helps users identify commits that incorporate user-collected fragments.

\section{Design Considerations}

Discovering and expanding new fragments is not a simple task but rather a compound process that requires iterative querying of known fragments \cite{sultanum2020understanding, choe2021papers101}.
We simulated the iterative process of exploring information fragments in Git repository based on the researches of literature search; 1) Starting from a handful of fragments, one iteratively searches target areas for a commit containing parts of fragments, a bunch of commits within a certain period or a release; 
2) One can discover new fragments while understanding the context of target areas by investigating their overview and details;
3) Then, the collected fragments are reused for the next query; 
4) Finally, one can identify the precise context and revisions incorporating the collected fragments. 

To support this iterative process of exploring information fragments, we derived design considerations:
\begin{itemize}[noitemsep]
    \item \textbf{C1}: \textbf{Narrow down the range of the exploration.} 
    Support filtering the \textit{target area} to narrow down the exploration scope. 

    \item \textbf{C2}: \textbf{Obtain an overview of the selection and examine details.}
    Gain an overview of numerous commits in the \textit{target area} and examine details on demand.
    
    \item \textbf{C3}: \textbf{Scoping topological neighbors or clusters.}
    Scoping the \textit{target area} to include topological neighbors or clusters of similar commits.
    
    \item \textbf{C4}: \textbf{Expand the collected fragments.}
    Expand the information fragments by discovering the residual fragments.
\end{itemize}
C1 and C2 come from the fact that software repositories are huge and complicated. 
C3 is to support the topological exploration of the repository. Still at the same time, we aim to provide users with serendipitous finding of new information fragments within topologically or temporally adjacent contexts.



 



\section{Interactive Visualization}
In this section, we describe the visualization and interaction design of \githruext{} for exploring information fragments in Git repository. 

\parahead{Scoping the exploration range}
The base system, Githru helps the initial scoping of the exploration range in the stem graph through temporal filtering and clustering by releases or similarity of commits (C1).
One can select the scope according to initial fragments of issued context such as author, issue date, release, and keywords. 
The scope is not necessarily small or specific because it contains topologically neighboring commits on the same branch; therefore one can discover new potentially relevant fragments within the topologically or temporally similar context (C3).
The clusters included in the selected scope are displayed at the top of the screen as shown in \autoref{fig:teaser}a, and they are encoded in the rectangular shape with the count of commits included in the bottom right corner as the graph representation of the base system.
We provide a vertical slider to control the granularity of clustering to help users narrow down the scope without going back and forth to the selection (C1).

\parahead{Dimension values overview}
One can examine the detail of commits in the clusters by clicking on each cluster (\autoref{fig:teaser}b).
However, a large number of commits in a cluster could make it hard to discover fragments. 
Hence, we depict the overview of dimension values in the dimension value table (\autoref{fig:teaser}c) aligned with clusters above (C2).
The leftmost column of the table displays the top five most frequent values of each dimension in the selected scope and the other columns also present top five values of each cluster vertically.
In case of the file and directory dimensions, the LOC (Lines of Code) value is also depicted to show the amount of changed codes.
The color of each value cell is derived from each dimension's color (as shown in the legend at the upper right corner), and its brightness level decreases in the order of frequent ranking.
We linked the cells of the same rank with curved lines to help users track changes in the rank of the frequent values; hence one can identify how the importance of the cell values changes.


\parahead{Inspection by fragments}
One can inspect whether each cluster incorporates known fragments to discover more fragments and identify the appropriate commits containing the desired context (C4).
By clicking on an information fragment in the dimension value table or the detail of commits, the inspection view describes whether each cluster contains the fragment (\autoref{fig:teaser}d). 
When inspecting multiple fragments, one can identify the appropriate clusters by checking the matched sum in the last row, representing how much the cluster is related to the fragments; 
This approach is for increasing the recall rather than the precision of search to discover potential fragments. 

\parahead{Fragment History and Pinning}
The fragment view presents their history when the fragments are clicked (\autoref{fig:teaser}e; C4). 
Also, one can collect discovered fragments by pinning. They can be reused for inspecting a new selection scope at once.







\section{Qualitative Study and Discussion}
We performed a case study with a developer (D1; 12 years of development experience) who had been using Git in practice. 
We demonstrated the tool with the interviewee and discussed its usability, effectiveness, limitations for deriving further works to be deployed in practice.

D1 was mainly interested in obtaining both an overview and details through the dimension value table.
He went through the overall values based on the leftmost column by comparing them to the clusters' values. 
He could understand whether or not to examine the clusters by the colors of the values.
He then investigated the detail of the clusters having common values and sometimes examined the detail of the cluster without common values (\autoref{fig:teaser}f) to check the reason why the cluster became the outlier in the selection scope.

He addressed the necessity of filtering by dimension in the case of a large number of value cells.
Moreover, he suggested the aggregated view of the information that matches the collected fragments.
He pointed out that it was confusing to distinguish the LOC and commit count value of the file/directory dimensions; we need to separate them using different color channels or a hierarchical view.

He also noted that the system needs to provide the detail of changed source codes.
Our system focused on Git metadata as information fragments but not on changed source codes due to concerns about performance.
To be deployed in practical, we need to extend the coverage of \githruext{} to support the code exploration in detail by attaching a search engine (i.e., OpenGrok\footnote{\vspace{-15pt} https://oracle.github.io/opengrok/.}) as the back-end system.

Currently, the exploration can be locally performed in the selection scope. 
However, in the case of solving a recurrent problem, it is necessary to examine large scale of past commits globally.
The scalable visualization using a t-SNE scatterplot to identify similar commits containing the information fragments of the problem can alleviate this issue.
Moreover, 
Also, conducting the performance comparison study to existing tools can confirm the effectiveness of our system in practice.
We leave these issues for future works.










\bibliographystyle{abbrv-doi}

\bibliography{main}

\begin{thebibliography}{1}

\bibitem{choe2021papers101}
K.~Choe, S.~Jung, S.~Park, H.~Hong, and J.~Seo.
\newblock Papers101: Supporting the discovery process in the literature review
  workflow for novice researchers.
\newblock In {\em 2021 IEEE Pacific Visualization Symposium (PacificVis)}.
  IEEE, 2021.

\bibitem{cosentino2015gitana}
V.~Cosentino, J.~L.~C. Izquierdo, and J.~Cabot.
\newblock Gitana: a sql-based git repository inspector.
\newblock In {\em International Conference on Conceptual Modeling}, pp.
  329--343. Springer, 2015.

\bibitem{feiner2018repovis}
J.~Feiner and K.~Andrews.
\newblock Repovis: Visual overviews and full-text search in software
  repositories.
\newblock In {\em 2018 IEEE Working Conference on Software Visualization
  (VISSOFT)}, pp. 1--11. IEEE, 2018.

\bibitem{fritz2010using}
T.~Fritz and G.~C. Murphy.
\newblock Using information fragments to answer the questions developers ask.
\newblock In {\em 2010 ACM/IEEE 32nd International Conference on Software
  Engineering}, vol.~1, pp. 175--184. IEEE, 2010.

\bibitem{kim2020githru}
Y.~Kim, J.~Kim, H.~Jeon, Y.-H. Kim, H.~Song, B.~Kim, and J.~Seo.
\newblock Githru: Visual analytics for understanding software development
  history through git metadata analysis.
\newblock {\em IEEE Transactions on Visualization and Computer Graphics}, 2020.

\bibitem{sultanum2020understanding}
N.~Sultanum, C.~Murad, and D.~Wigdor.
\newblock Understanding and supporting academic literature review workflows
  with litsense.
\newblock In {\em Proceedings of the International Conference on Advanced
  Visual Interfaces}, pp. 1--5, 2020.

\end{thebibliography}
\end{document}